\documentstyle[12pt]{article}
\begin{document}
\thispagestyle{empty}
\setcounter{page}0

~\vfill
\begin{center}
{\Large\bf	New interactions in the radiative pion decay} \vfill

{\large	M. V. Chizhov}
 \vspace{1cm}

{\em Center of Space Research and Technologies, Faculty of Physics,
University of Sofia, 1164 Sofia, Bulgaria}

\end{center}  \vfill

\begin{abstract}

The preliminary results of PIBETA experiment strongly
suggest the presence of non $V\!-\!A$ anomalous 
interactions in the radiative
pion decay. Without a guiding idea about the nature of these new
interactions it is very difficult to interpret and fit
experimental data. We assume that they arise as a result of 
the exchange of new intermediate chiral bosons which interact
anomalously with matter. Their mixing with $W$-bosons leads 
effectively to a small anomalous weak moment for $W$.
Based on these assumptions, we show that the most general
form of the radiative pion decay rate 
can be parametrized by three new coupling constants
of the anomalous interactions. 

\end{abstract}

\vfill

\newpage

\section{Introduction}

The Standard Model (SM) is now a well established theory of the electroweak
and the strong interactions. Its experimental success is impressively great
and, therefore, all anomalies in the present and the past experiments 
are usually considered as their artifacts or statistical fluctuations.
Nevertheless, from the general point of view we believe that SM is just 
the limit of a more general theory with new particles and 
an enlarged symmetry. 

Many people trust that this is the supersymmetry (SUSY). 
The present phenomenological SUSY models
are constructed by simply doubling of the known particles without
introducing really new kinds of fields. Therefore, for their description 
the contemporary framework of the quantum field theory can be used. 
In such models all processes can be computed and predictions for their 
experimental manifestation can be made.

In the present paper we discuss a new kind of vector particles, 
whose full theory is not constructed yet. 
However, a phenomenological approach is possible, 
and, moreover, it is motivated by the available experimental data.
The introduction of such type of particles is natural 
and does not contradict the basic axioms of the quantum field theory.
In the paper~\cite{NJL} we have shown that such type of particles
exists as composite quark-antiquark hadron vector resonances.
Based on this idea, the dynamic properties of the vector mesons and
the new mass relations among the hadron vector resonances have been derived.

The preliminary results of the PIBETA experiment~\cite{PIBETA} 
on the radiative pion decay (RPD) may represent
one of the main indications of the existence of such particles 
also on a fundamental level.  Using a simple
phenomenological model we will show how these results can be interpreted.

\section{The tensor interactions}

The radiative pion decay $\pi\to\nu e\gamma$
is a unique system for searching physics beyond SM.
Due to chirality suppression the usual electroweak interactions are very weak
while new interactions, which do not preserve chirality of fermions, can be
enhanced. However, the enhancement occurs only in a certain part 
of the kinematically allowed region and, therefore,
the previous experiments~\cite{old} have not detected the new interactions. 

The first experiment, which has alarmed about deviations from SM in RPD, was
ISTRA experiment~\cite{Bolotov}. The RPD in flight was investigated 
in a wide kinematical region with dramatic conclusions for SM. 
It has been found that the measured number of events is about 30\% smaller 
than the expected ones.
This cannot be explained neither by radiative corrections~\cite{Nikitin}
nor by SUSY extension of SM~\cite{SUSY}. The interpretation of these results
in the framework of SM leads to a huge violation of CVC hypothesis.

In order to describe this strange result a new interaction with a tensor
lepton current
\begin{equation}
M_T = -\frac{eG}{\sqrt{2}} F_T~ \varepsilon_\alpha(q) q_\beta~
\bar{e}\sigma_{\alpha\beta}(1-\gamma^5)\nu
\label{MT1}
\end{equation}
with $F_T = -(5.6\pm 1.7)\times 10^{-3}$ was introduced~\cite{Pob}.
Here $G=G_F V_{ud}$ and $\varepsilon_\alpha(q)$ is the photon
polarization vector.
The additional interaction helps to explain the lack of events with respect 
to SM prediction, caused by 
the destructive interference of the new term with inner-bremsstrahlung (IB)
amplitude. Moreover, it leads to a correct events distribution over 
the Dalitz plot.

Such type of a phenomenologically introduced amplitude can be induced
by a four-fermion tensor interaction of quark and lepton currents
\begin{equation}
{\cal L}_T = -\frac{G}{\sqrt{2}} f_T~ 
\bar{u}\sigma_{\alpha\beta}(1-\gamma^5)d~
\bar{e}\sigma_{\alpha\beta}(1-\gamma^5)\nu.
\label{LT1}
\end{equation}
As far as tensor intermediate bosons are absent in SM and its popular
extensions, the conclusion was made that such type of interaction can be
generated only by a leptoquark exchange~\cite{Herczeg}. However, according
to the Grand Unified Models (GUT) the leptoquarks should be very massive
$\sim 10^{15}$ GeV and, hence, their influence at the electroweak scale 
should be negligibly small.

The introduction of the new tensor interaction (\ref{MT1})
leads to another problem,
following from the strong constraint on $F_T$~\cite{Vol} from pion
decay $\pi\to e\nu$. On one hand the tensor interaction does not contribute 
directly to semileptonic two-particle pion decay $\pi\to e\nu$, due to
kinematical reasons. 
On the other hand, however, owing to electromagnetic radiative
corrections, the pseudotensor current 
$\bar{u}\sigma_{\alpha\beta}\gamma^5d$ leads to
a generation of a pseudoscalar quark current, to which the pion decay is 
very sensitive. As a result, $F_T$ should be two orders of magnitude smaller
than its required value for the explanation of ISTRA results.

To avoid this problem a new tensor amplitude
\begin{equation}
M'_T = -\frac{eG}{\sqrt{2}}F'_T~\left[q_\alpha\varepsilon_\lambda(q)-
q_\lambda\varepsilon_\alpha(q)\right]
\frac{Q_\lambda Q_\beta}{Q^2}~
\bar{e}\sigma_{\alpha\beta}(1-\gamma^5)\nu,
\label{MT2}
\end{equation}
was introduced~\cite{MPL} in addition to (\ref{MT1}).
Here $Q_\alpha=(p-q)_\alpha$ is the momentum transfer to the lepton pair.
This amplitude is generated by the following four-fermion tensor interaction
\begin{equation}
{\cal L}'_T =  -\frac{4G}{\sqrt{2}}f'_T~ 
\bar{u}\sigma_{\alpha\lambda}(1+\gamma^5)d~
\frac{Q_\lambda Q_\beta}{Q^2}~
\bar{e}\sigma_{\alpha\beta}(1-\gamma^5)\nu.
\label{LT2}
\end{equation}
Since this interaction includes quark current with 
an opposite chirality to the one in (\ref{LT1}), 
and also the interaction (\ref{LT1}) can be identically rewritten as
\begin{equation}
\bar{u}\sigma_{\alpha\beta}(1-\gamma^5)d~
\bar{e}\sigma_{\alpha\beta}(1-\gamma^5)\nu\equiv 4~
\bar{u}\sigma_{\alpha\lambda}(1-\gamma^5)d~\frac{Q_\lambda Q_\beta}{Q^2}~
\bar{e}\sigma_{\alpha\beta}(1-\gamma^5)\nu,
\end{equation}
then the pseudotensor terms $\bar{u}\sigma_{\alpha\beta}\gamma^5d$ 
cancel out in the sum of the eqs.
(\ref{LT1}) and (\ref{LT2}), if $f_T=f'_T$. The tensor term
$\bar{u}\sigma_{\alpha\beta}d$ does not contribute to pseudoscalar pion decay 
due to parity conservation in electromagnetic interactions.

Therefore, the fit of experimental data with only one amplitude (\ref{MT1})
is not acceptable because of the contradiction with present experimental data
on pion decay. Consequently, it is not a surprise that the values $F_T$ for
ISTRA and PIBETA experiments, which run in different kinematical
regions, are different.

\section{The model}

The sum of the interactions (\ref{LT1}) and (\ref{LT2}) 
is not the most general Lagrangian including different tensor currents.
Following the idea of renormalizability one can generate the effective
four-fermion tensor interactions (\ref{LT1}) and (\ref{LT2}) by exchange
of new chiral vector particles, which interact with matter only anomalously
$g_T T^\pm_\alpha \hat{Q}_\beta~
\bar{\psi}\sigma_{\alpha\beta}(1\pm\gamma^5)\psi$,
where $\hat{Q}_\alpha\equiv Q_\alpha/\sqrt{Q^2}$.
If these particles
mix with $W$-bosons, the latter could acquire the anomalous weak moment
introduced in~\cite{hep}.
Hence, the most general interaction of $W$-boson can be written in the form 
\begin{eqnarray}
{\cal L}_{int}=&&\hspace{-0.5cm}-\frac{1}{2\sqrt{2}}W^-_\alpha\left[
g~\bar{e}\gamma_\alpha(1-\gamma^5)\nu
+ig^T_e~\hat{Q}_\beta~\bar{e}\sigma_{\alpha\beta}(1-\gamma^5)\nu\right]
\nonumber\\
&&\hspace{-0.5cm}-\frac{1}{2\sqrt{2}}W^+_\alpha\left[
g~\bar{u}\gamma_\alpha(1-\gamma^5)d
-ig^T_V~\hat{Q}_\beta~\bar{u}\sigma_{\alpha\beta}d
-ig^T_A~\hat{Q}_\beta~\bar{u}\sigma_{\alpha\beta}\gamma^5d\right]
+{\rm h.c.},
\label{int}
\end{eqnarray}
introducing in general three new independent parameters $g^T_e$, $g^T_V$
and $g^T_A$. 
It leads effectively to five new four-fermion interactions 
\begin{eqnarray}
{\cal L}_{eff}=-\frac{G}{\sqrt{2}}&&\hspace{-0.5cm}\left[
\bar{u}\gamma_\alpha(1-\gamma^5)d
-i\hat{Q}_\lambda~\bar{u}\sigma_{\alpha\lambda}(g^T_V+g^T_A\gamma^5)d\right]
\bar{e}\gamma_\alpha(1-\gamma^5)\nu
\nonumber\\
-\frac{ig^T_eG}{\sqrt{2}}&&\hspace{-0.5cm}\left[
\bar{u}\gamma_\alpha(1-\gamma^5)d
-i\hat{Q}_\lambda~\bar{u}\sigma_{\alpha\lambda}(g^T_V+g^T_A\gamma^5)d\right]
\hat{Q}_\beta~\bar{e}\sigma_{\alpha\beta}(1-\gamma^5)\nu
\label{eff}
\end{eqnarray}
with only three independent parameters. In order to get the amplitude
of RPD it is necessary to calculate the following matrix elements for
$\pi$-$\gamma$ transition
\begin{eqnarray}
\langle\gamma(q)|\bar{u}\gamma_\alpha(1-\gamma^5)d|\pi(p)\rangle\!\!&=&\!\!
-\frac{e}{m_\pi}\varepsilon_\beta(q)\left\{
F^0_V\epsilon_{\alpha\beta\rho\sigma}p_\rho\ q_\sigma+iF^0_A\left[
(pq)g_{\alpha\beta}-q_\alpha p_\beta\right]\right\}
\\
\langle\gamma(q)|\bar{u}\sigma_{\alpha\beta}\gamma^5 d|\pi(p)\rangle\!\!&=&\!\!
-eF^0_T\left[q_\alpha\varepsilon_\beta(q)
-q_\beta\varepsilon_\alpha(q)\right], 
\end{eqnarray}
which can be parametrized by three form factors $F^0_V$, $F^0_A$ and
$F^0_T$. These form factors weakly depend on the square of momentum 
transfer to the lepton pair $Q^2=(p-q)^2$ and can be taken as constants.
 
Assuming CVC hypothesis,
the vector form factor $F^0_V$ is directly related to the 
$\pi^0\to\gamma\gamma$ amplitude and can be extracted from the experimental
width of this decay
\begin{equation}
F^0_V=\frac{1}{\alpha}\sqrt{
\frac{2\Gamma(\pi^0\to\gamma\gamma)}{\pi m_{\pi^0}}}=
0.0262\pm 0.0009
\end{equation}
This value is in fair agreement with the calculations in the relativistic
quark model (RQM) and with
the leading order calculations of the chiral perturbation theory (CHPT)
\begin{equation}
F^0_V=\frac{1}{4\pi^2}\frac{m_\pi}{F_\pi}\approx 0.0270,
\end{equation}
where $F_\pi=(130.7\pm 0.4)$~MeV is the pion decay constant.

An axial form factor $F^0_A\simeq 0.4 F^0_V$ has been measured 
in previous experiments~\cite{old} in the kinematical region 
where the contribution of new tensor terms is not essential. 
This value is also in agreement with the calculations in CHPT~\cite{CHPT}. 

The tensor form factor $F^0_T$ can be calculated by applying the QCD 
sum rules techniques~\cite{Ian} and the PCAC hypothesis
\begin{equation}
F^0_T=\frac{1}{3}\chi\frac{\langle 0|\bar{q}q|0\rangle}{F_\pi}
\approx 0.2
\end{equation}
here $\chi=-(5.7\pm 0.6)$~GeV$^{-2}$ is the magnetic susceptibility 
of the quark condensate and its vacuum expectation value is 
${\langle 0 |\bar{q}q|0\rangle}\approx-(0.24~{\rm GeV})^3$.

The most general matrix element for radiative pion decay 
\begin{equation}
M=M_{IB}+M_{SD}+M_T+M'_T
\end{equation}
can be rewritten through four form factors
$F_V$, $F_A$, $F_T$ and $F'_T$~\cite{Pob03}.
Here the first term
\begin{equation}
M_{IB}=-i\frac{eG}{\sqrt{2}}F_\pi m_e\varepsilon_\alpha(q)
\bar{e}\left[\left(\frac{k_\alpha}{kq}-\frac{p_\alpha}{pq}\right)
-\frac{i\sigma_{\alpha\beta}q_\beta}{2kq}\right](1-\gamma^5)\nu
\end{equation}
is a QED radiative correction to the $\pi\to e\nu$ decay (IB) and
it does not depend on any form factors. The structure-dependent
amplitude
\begin{equation}
M_{SD}=\frac{eG}{\sqrt{2}m_\pi}\varepsilon_\beta(q)\left\{
F_V\epsilon_{\alpha\beta\rho\sigma}p_\rho q_\sigma+iF_A\left[
(pq)g_{\alpha\beta}-p_\alpha q_\beta\right]\right\}
\bar{e}\gamma_\alpha(1-\gamma^5)\nu
\end{equation}
is parametrized by two form factors $F_V$ and $F_A$.
However, due to mixed tensor-vector interactions (second term
in (\ref{eff})) the form factors $F_V$ and $F_A$ are not constants anymore
and depend on $Q^2$ in a specific way
\begin{eqnarray}
F_V\!\!&=&\!\!F^0_V+\frac{m_\pi}{\sqrt{Q^2}}~g^T_V F^0_T,
\nonumber\\
F_A\!\!&=&\!\!F^0_A+\frac{m_\pi}{\sqrt{Q^2}}~g^T_A F^0_T.
\label{FVA}
\end{eqnarray}
The expressions for the tensor amplitudes $M_T$ and $M'_T$ coincide with  
(\ref{MT1}) and (\ref{MT2}), but the form factors $F_T$ and $F'_T$
are not constants. In our model they have strong $Q^2$ dependence
and due to vector-tensor mixed interactions (third term in (\ref{eff})) they
depend also on the vector $F^0_V$ and the axial-vector $F^0_A$
form factors as well
\begin{eqnarray}
F_T\!\!&=&\!\!-g^T_e\left[
g^T_V F^0_T+\frac{\sqrt{Q^2}}{m_\pi}~ F^0_V\right],
\nonumber\\
F'_T\!\!&=&\!\!-g^T_e\left[
\left(g^T_V+g^T_A\right) F^0_T+
\frac{\sqrt{Q^2}}{m_\pi}~ \left(F^0_V+F^0_A\right)\right].
\label{FT}
\end{eqnarray}

\section{The decay rate}

In the general case the decay rate depends on the squares of the amplitudes
and their various interference terms. In our case,
when the electron mass is vanishingly small,
the interference between the SD amplitude on the one hand,
and the IB amplitude or the tensor amplitudes on the other hand, can be 
safely neglected.

In this approximation the differential decay width is
\begin{eqnarray}
{{\rm d}^2 \Gamma\over{\rm d}x{\rm d}\lambda}=
\frac{\alpha}{2\pi}\Gamma_{\pi\to e\nu}\left\{
IB(x,\lambda)+a^2_{SD}\left[(F_V+F_A)^2SD^+(x,\lambda)
+(F_V-F_A)^2SD^-(x,\lambda)\right]\right.
\nonumber\\
\left.+2 a_{SD} \left[2(F_T-F'_T)+F'_T x\right]I(x,\lambda)
+2 a^2_{SD}\left[2F_T(F_T-F'_T)+F'^2_T\right]T(x,\lambda)
\right\}
\label{rate}
\end{eqnarray}
where $a_{SD}=m^2_\pi/2F_\pi m_e\approx 145.8$,
\begin{eqnarray}
IB(x,\lambda)=\frac{(1-x)^2+1}{x}\frac{1-\lambda}{\lambda},\hspace{0.5cm}
SD^+(x,\lambda)=(1-x)x^3\lambda^2,\hspace{2.9cm}
\nonumber\\
SD^-(x,\lambda)=(1-x)x^3(1-\lambda)^2,\hspace{0.3cm}
I(x,\lambda)=x(1-\lambda),\hspace{0.3cm}
T(x,\lambda)=x^3(1-\lambda)\lambda.
\label{rho}
\end{eqnarray}
In the pion rest frame the variables $x$ and $\lambda$ are defined as
$x=2E_\gamma/m_\pi$ and $\lambda=2E_e/m_\pi\sin^2(\theta_{e\gamma}/2)$.

To analyse the contribution of the new terms into the decay rate (\ref{rate})
one can use the analytical expressions for 
the Dalitz plot densities (\ref{rho}) and the form factors dependences
(\ref{FVA}) and (\ref{FT}).
The squared momentum transfer to the lepton pair
$Q^2=m^2_\pi(1-x)\ge m^2_e$ is constrained from below and
the maximal effects of the new terms on $F_V$ and $F_A$
reveal exactly at this low limit. In other words, the events with 
the maximal photon energy $x\simeq 1$ represent the interesting
region of the Dalitz plot, where the new terms effect is maximal.
To avoid the contradiction with the discussed
constraint from the pion decay $\pi\to e\nu$, the anomalous axial
coupling constant $g^T_A$ should be very small. Hence, for fitting 
one can use actually only two of the three independent parameters. 

In the case when the anomalous tensor coupling constant $g^T_V$
is negative, one can get effectively lower value for the vector form factor
$F_V$ than the CVC predicted one. Noticeably, 
exactly such effect has been observed in
ISTRA~\cite{Bolotov} and PIBETA~\cite{web} fits. 

Another manifestation
of the new interactions can be observed as a destructive interference
between the IB amplitude and the amplitudes with the tensor lepton currents
at $x\simeq 1$. This result immediately follows from the expressions
(\ref{FT}) for the tensor form factors, in case the natural universality 
hypothesis about the sign of the tensor coupling constants 
$g^T_e$ and $g^T_V$ is accepted. The effect of the destructive interference 
is the maximal one at $\lambda\simeq 0$.

\section{Conclusions}

In this paper we present a simple model for an explanation of ISTRA
and PIBETA results on the radiative pion decay. A small admixture of
anomalous couplings for the $W$-boson interactions with a specific
form (\ref{int}) allows to explain the essential features of the observed
deviations from SM in these experiments. Hopefully, the correct fitting
according to this model of experimental data of high statistics PIBETA 
experiment will reveal new directions for
probing physics beyond the Standard Model.

\section*{Acknowledgements}

I am grateful to A. Olshevsky, V. G. Kadyshevsky, A. Dorokhov,
A. Poblaguev, S.~M.~Korenchenko, D. Mzhavia, N. Kuchinsky, D. Po\v{c}ani\'c, 
E. Frle\v{z}, and W. Bertl for useful discussions and overall help.

I acknowledge the hospitality of the Joint Institute for Nuclear Research
and Paul Scherrer Institute, where this work has been fulfilled.
I highly appreciate the financial support of JINR for my presentation
of this work at the Dzhelepov Laboratory of Nuclear Problems,
JINR and at the PIBETA Collaboration Meeting, PSI.

\pagebreak[3]

\end{document}